\documentstyle[aps,floats,epsfig]{revtex}

\def\pslash{\rlap{\hspace{0.02cm}/}{p}}
\def\nslash{\rlap{\hspace{0.02cm}/}{n}}

\begin{document}
\draft
\twocolumn[\hsize\textwidth\columnwidth\hsize\csname@twocolumnfalse%
\endcsname


\title{\boldmath
Unparticle Physics with Jets
\unboldmath}

\author{Matthias Neubert
\\[0.1cm]
{\em Institut f\"ur Physik (ThEP), Johannes Gutenberg-Universit\"at\\ 
D--55099 Mainz, Germany}\\[0.1cm]
and\\[0.1cm]
{\em Fermi National Accelerator Laboratory\\
P.O. Box 500, Batavia, IL 60510, U.S.A.}}
\maketitle

\begin{abstract}
Using methods of effective field theory, we show that after resummation of Sudakov logarithms the spectral densities of interacting quark and gluon fields in ordinary quantum field theories such as QCD are virtually indistinguishable from those of ``unparticles" of a hypothetical conformal sector coupled to the Standard Model, recently studied by Georgi. Unparticles are therefore less exotic that originally thought. Models in which a hidden sector weakly coupled to the Standard Model contains a QCD-like theory, which confines at some scale much below the characteristic energy of a given process, can give rise to signatures closely resembling those from unparticles.
\end{abstract}

\pacs{Preprint: MZ-TH/07-13, FERMILAB-PUB-07-402-T}]
\narrowtext

\section{Introduction}

In recent work \cite{Georgi:2007ek}, Georgi has introduced the notion of ``unparticles" into elementary-particle physics and speculated that they might lead to spectacular signals at future colliders such as the LHC. He envisions a scheme in which at very high energy Standard Model fields can interact with the fields of some hidden sector via the exchange of heavy messenger particles with masses $M_{\cal U}$. At lower energies the interactions between the two sectors are described by nonrenormalizable operators suppressed by powers of $M_{\cal U}$. He then assumes that the hidden-sector theory has a non-trivial infrared fixed point, so that scale invariance emerges at some scale $\Lambda_{\cal U}$. In the effective theory below this scale the hidden-sector operators $O_{\rm UV}$ match onto ``unparticle operators" $O_{\cal U}$ of an effective theory with conformal symmetry. The couplings of these operators to Standard Model operators have the generic form
\begin{equation}\label{genericint}
   \frac{C_{\cal U}\,\Lambda_{\cal U}^{d_{\rm UV}-d_{\cal U}}}%
        {M_{\cal U}^{d_{\rm UV}+d_{\rm SM}-4}}\,
        O_{\rm SM}\,O_{\cal U} \,,
\end{equation}
where $d_{\rm UV}$ is the scaling dimension of the original operator in the hidden sector, $d_{\cal U}$ is the scaling dimension of the unparticle operator $O_{\cal U}$, and $C_{\cal U}$ is a dimensionless Wilson coefficient. At low energies the unparticle operators with the lowest dimension will give rise to the leading effects. In a strongly interacting conformal theory, there is no reason why this dimension should be an integer.

In order to calculate the probability distribution for processes in which unparticles are produced one needs the density of unparticle states, which can be defined via the Fourier transform of the two-point function $\langle 0|\,O_{\cal U}(x)\,O_{\cal U}^\dagger(0)\,|0\rangle$. The dependence of this correlator on $x^2$, and hence the  dependence of the spectral density on $p^2$, is fixed by conformal invariance. The spectral density associated with an unparticle operator of scaling dimenson $d_{\cal U}$ is \cite{Georgi:2007ek}
\begin{equation}\label{rho}
   \rho(p^2) = \frac{N_\eta}{\Gamma(\eta)} \left(p^2\right)^{\eta-1}
    \,; \qquad 
   \eta = d_{\cal U}-1 \,,
\end{equation}
where for brevity we omit $\theta$-functions ensuring that $p^0\ge 0$ and $p^2\ge 0$. The conformal dimension of the operator $O_{\cal U}$ is assumed to satisfy $1<d_{\cal U}<2$, so that $0<\eta<1$. The choice of the normalization constant $N_\eta$ is a matter of convenience and is irrelevant to our discussion. Note, however, that $N_\eta$ has scaling dimension $-2\eta$, ensuring that the spectral density has scaling dimension $-2$. We will furthermore assume that $N_0=1$.

In the limit $\eta\to 0$ (i.e., $d_{\cal U}\to 1$) the spectral density reduces to that of a free particle, $\rho(p^2)=\delta(p^2)$. The parameter $\eta$ characterizes the extent to which unparticles differ from free particles. In \cite{Georgi:2007ek}, it was argued that the spectral density (\ref{rho}) provides a generalization of $n$-particle phase space to the case where $n=d_{\cal U}$ is not an integer. The author went on to state that the discovery of ``unparticle stuff" with such nontrivial scaling behavior would be ``a much more striking discovery than the more talked about possibilities of SUSY or extra dimensions". Subsequently, phenomenological implications of the peculiar form (\ref{rho}) of the unparticle spectral density have been explored in a large number of publications \cite{Georgi:2007si}--\cite{Delgado:2007dx}. A weakness of this branch of phenomenology is that to date no explicit model has been constructed that would exhibit unparticle behavior. It is therefore unknown in which way unparticles couple to ordinary particles and if they carry Standard Model gauge interactions. 

However, if unparticles couple to the Standard Model and hence may give rise to observable effects, then the conformal symmetry in the unparticle sector is broken by Standard Model loops. Unavoidably this will lead to a modification of the spectral density (\ref{rho}). The precise form of the resulting scaling violations cannot be predicted without a concrete realization of the unparticle scenario. A particularly dangerous source of conformal-symmetry breaking arises if scalar unparticles couple to the Standard Model Higgs field, as for $d_{\cal U}<2$ this leads to a relevant operator in the low-energy theory \cite{Fox:2007sy} (see also \cite{Schabinger:2005ei,Strassler:2006im,Patt:2006fw}). The presence of such an operator would lead to severe constraints, which most likely would render unparticle effects invisible at present energies \cite{Bander:2007nd}, since it would imply a breaking of conformal invariance at the electroweak scale. A simple model of a less severe kind of conformal symmetry breaking is obtained by introducing a mass gap \cite{Fox:2007sy},
\begin{equation}\label{rhom}
   \rho(p^2,m^2) = \frac{N_\eta}{\Gamma(\eta)} 
   \left(p^2-m^2\right)^{\eta-1} \,,
\end{equation}
where now $p^2\ge m^2$ and $p^0\ge 0$. For $\eta\to 0$ this reduces to the spectral density of a free massive particle, $\rho(p^2,m^2)=\delta(p^2-m^2)$. Note that while the form of (\ref{rho}) is determined by the requirement of conformal invariance in the unparticle sector, relation (\ref{rhom}) is but a simple model for the spectral density in a more complicated theory in which conformal invariance is broken. In general, it follows that a scenario such as the one envisioned in \cite{Georgi:2007ek} can only be realized in a ``conformal window" below the scale $\Lambda_{\cal U}$ and above a scale $\Lambda_{\rm CSB}$ characterizing conformal symmetry breaking. Only for energies inside this window and much above $\Lambda_{\rm CSB}$ the scaling behavior (\ref{rho}) of the unparticle spectral density can manifest itself in a characteristic scaling of cross sections or decay rates. 
Unfortunately, this important role of conformal symmetry breaking has been largely ignored in the literature on unparticles. In the examples above, we expect $\Lambda_{\rm CSB}\sim\Lambda_{\rm EWSB}$ if conformal symmetry is broken by the presence of a relevant coupling of unparticles to the Higgs sector, and $\Lambda_{\rm CSB}\sim m$ is it is broken by a mass gap. Even inside the conformal window, the scaling behavior resulting from the spectral density (\ref{rho}) will receive corrections from Standard Model loop effects. 

Following the work \cite{Georgi:2007ek}, several authors have tried to demystify the notion of unparticles by relating them to systems of ordinary particles. Interpolating the continuous unparticle spectral density (\ref{rho}) by an infinite sum over densly spaced $\delta$-functions, an unparticle can be represented by an infinite tower of massive particles with mass-dependent decay constants \cite{Stephanov:2007ry},
\begin{equation}
   \rho(p^2) = \sum_n F_n^2\,\delta(p^2-M_n^2) \,.
\end{equation}
This ``deconstruction" of the unparticle can, of course, be performed for any function $\rho(p^2)$ irrespective of whether it exhibits the particular form (\ref{rho}) dictated by conformal symmetry. If the spacing of the massive particles is chosen to be equidistant, then it is tempting to interpret them as the Kaluza-Klein tower of a fundamental field propagating in an extra dimension. For instance, the Kaluza-Klein tower of a massless scalar field propagating in a flat extra dimension provides a discretization of the unparticle spectral density with $\eta=1/2$ (i.e., $d_{\cal U}=3/2$) \cite{Cheung:2007ap}. In the case of warped extra dimensions, the conformal dimension $d_{\cal U}$ is linked to the mass of a bulk scalar field, so that a whole class of unparticle models can be viewed as holographic duals of Randall-Sundrum models \cite{Stephanov:2007ry}.

In a similar spirit, unparticles can be obtained from a special limit of higher-dimensional models, in which the Standard Model is extended by singlet fields living in extra dimensions \cite{van der Bij:2007um}. As a consequence, Standard Model fields that normally contain single-particle peaks satisfy more general spectral representations. Unparticles can also be interpreted as a particular case of fields with continuously distributed mass \cite{Krasnikov:2007fs}.

\section{Sudakov resummation for jets}

We demonstrate in this Letter that the spectral densities (\ref{rho}) and (\ref{rhom}) are not as unusual as claimed in \cite{Georgi:2007ek}. It is well known that the full propagators of interacting particles in quantum field theory obey a K\"all\'en-Lehmann spectral representation, in which the spectral density differs from the simple-pole form valid for a free particle \cite{Peskin:1995ev}. This phenomenon has been well studied in perturbation theory. For instance, the Borel resummation of fermion-loop insertions on gauge-boson propagators in QED and QCD has been used to explore the asymptotic behavior of the perturbation series in gauge theories \cite{Beneke:1998ui}. The Borel-resummed gauge-boson propagator has a spectral density $\rho(p^2)\propto(p^2)^{-1-u}$, where $u$ is the Borel parameter conjugate to $\beta_0\alpha_s/4\pi$. The structure of the perturbative expansion is determined by the region near the origin in the Borel plane, so that the modification of the free propagator is a small effect. 

Significant modifications of the free propagator arise in (rather generic) situations in which several widely separated scales are present. In order to account for multiple emissions of soft and collinear radiation, the propagation of interacting quark and gluon fields in a theory such as QCD is described by jet functions (see, e.g., \cite{Sterman:1986aj,Catani:1989ne}). Considering the case of a massless quark for example, we define the full propagator by \cite{Becher:2006qw}
\begin{eqnarray}\label{jetfun}
   &&\hspace{-0.5cm}
    \langle 0|\,T\left[ \psi(x)\,\bar\psi(0) \right]
    |0\rangle_{\rm LCG} \nonumber\\
   &=& \int\frac{d^4p}{(2\pi)^4}\,e^{-ip\cdot x}
    \left[ \pslash\,{\cal J}(p^2,\mu) + \dots \right] .
\end{eqnarray}
This quantity is gauge dependent, and we define it using the light-cone gauge $n\cdot A=0$. A gauge-invariant definition could be obtained by multiplying the quark fields with Wilson lines. The dots in (\ref{jetfun}) represent terms proportional to $\nslash$, which we will ignore. We then define the spectral density as the discontinuity of the full propagator, 
\begin{equation}\label{jetfun1}
   \rho(p^2,\mu) 
   = \frac{1}{\pi}\,{\rm Im}\left[{i\cal J}(p^2,\mu)\right] .
\end{equation}
In the literature on perturbative QCD and soft-collinear effective theory this function is often called the jet function and denoted by $J(p^2,\mu)$, but we will continue to call it $\rho(p^2,\mu)$ for the purposes of this Letter. The jet function has support for $p^2\ge 0$ and $p^0\ge 0$. It is the analog of the spectral density (\ref{rho}) for the case of ordinary interacting particles.

At lowest order in perturbation theory the spectral density is $\rho(p^2)=\delta(p^2)$, corresponding to a free, massless particle. In higher orders logarithmic corrections appear. At first order in $\alpha_s$ one encounters terms of the form $\ln^n(p^2/\mu^2)/p^2$ with $n=0,1,2$ \cite{Manohar:2003vb,Bauer:2003pi,Bosch:2004th}. Similarly, at order $\alpha_s^n$ there appear up to $2n$ powers of logarithms. These Sudakov double logarithms arise due to the combined effects of soft and collinear gluon emissions. 

In physical processes involving jets there typically exist several widely separated energy scales. In particular, the invariant mass squared of a partonic jet can vary between some hadronic scale $\mu_0^2\sim\Lambda_{\rm QCD}^2$ up to some maximum value $M^2\equiv(p^2)_{\rm max}$ set by kinematics. In the case of deep-inelastic scattering at large $x$, for example, the kinematic range for the invariant mass of the final-state quark jet is $0\le p^2\le Q^2\,\frac{1-x}{x}$, where $Q^2$ is the hard momentum transfer and $x$ the Bjorken scaling variable. In such a case physical cross section are sensitive to large Sudakov logarithms, which must be resummed to all orders in perturbation theory. This is done by factorizing the cross section into different subprocesses and resumming the large logarithms by solving evolution equations (see e.g.\ \cite{Sterman:1986aj,Catani:1989ne,Manohar:2003vb,Becher:2006nr,Becher:2006mr}). In this case jet functions such as ${\cal J}(p^2,\mu)$ and $\rho(p^2,\mu)$ must be evolved from a high scale of order $M$ down to much lower scales. As we will now discuss, this can have a profound impact on their momentum dependence.

The spectral density $\rho(p^2,\mu)$ defined in (\ref{jetfun1}) obeys the renormalization-group evolution equation \cite{Becher:2006qw}
\begin{eqnarray}\label{Jrge}
   \frac{d\rho(p^2,\mu)}{d\ln\mu}
   &=& - \left[ 2\Gamma_{\rm cusp}(\mu)\,\ln\frac{p^2}{\mu^2}
    + 2\gamma^J(\mu) \right] \rho(p^2,\mu) \nonumber\\
   &&\hspace{-1.3cm}
    \mbox{}- 2\Gamma_{\rm cusp}(\mu) \int_0^{p^2}\!
    dp^{\prime 2}\,
    \frac{\rho(p^{\prime 2},\mu)-\rho(p^2,\mu)}{p^2-p^{\prime 2}} \,.
\end{eqnarray} 
The quantities $\Gamma_{\rm cusp}$ and $\gamma^J$ are anomalous dimensions, which depend on the renormalization scale only through the running coupling $\alpha_s(\mu)$. Their perturbative expansions are known to three-loop order. In particular, $\Gamma_{\rm cusp}$ is the cusp anomalous dimension of Wilson loops with light-like segments \cite{Korchemskaya:1992je}, which plays a central role in the physics of soft-gluon interactions (see e.g.\ \cite{Mert Aybat:2006wq}).
We stress that the form of the evolution kernel in (\ref{Jrge}) is exact; its simplicity is a consequence of dimensonal analysis combined with some magic properties of Wilson lines.

The exact solution to the evolution equation was obtained in \cite{Becher:2006nr}. It can be written in the form
\begin{eqnarray}\label{sonice}
   \rho(p^2,\mu_0) 
   &=& N(M,\mu_0) \left( p^2 \right)^{\eta-1} \nonumber\\
   &\times& \widetilde j\Big(\ln\frac{p^2}{M^2}+\partial_\eta,M\Big)\,
   \frac{e^{-\gamma_E\eta}}{\Gamma(\eta)} \,,
\end{eqnarray}
where $\partial_\eta$ denotes a derivative with respect to the quantity $\eta$, which is then identified with
\begin{equation}\label{eta}
   \eta = \int_{\mu_0^2}^{M^2}\!\frac{d\nu^2}{\nu^2}\,
   \Gamma_{\rm cusp}(\nu) \,.
\end{equation}
The normalization factor $N$ has scaling dimension $-2\eta$ and is given by
\begin{equation}
   \ln N(M,\mu_0) 
   = \int_{\mu_0^2}^{M^2}\!\frac{d\nu^2}{\nu^2}
   \left[ \Gamma_{\rm cusp}(\nu)\,\ln\frac{1}{\nu^2} + \gamma^J(\nu) \right] .
\end{equation}
This quantity is momentum-independent and will thus be irrelevant to our discussion. The function $\widetilde j(x,M)$ has a perturbative expansion free of large logarithms. At one-loop order \cite{Neubert:2004dd}
\begin{equation}
   \widetilde j(x,M) = 1 + \frac{C_F\alpha_s(M)}{4\pi}
   \left( 2x^2 - 3x + 7 - \frac{2\pi^2}{3} \right) ,
\end{equation}
while the two-loop expression can be found in \cite{Becher:2006qw}.

\begin{figure}
\begin{center}
\includegraphics[width=0.48\textwidth]{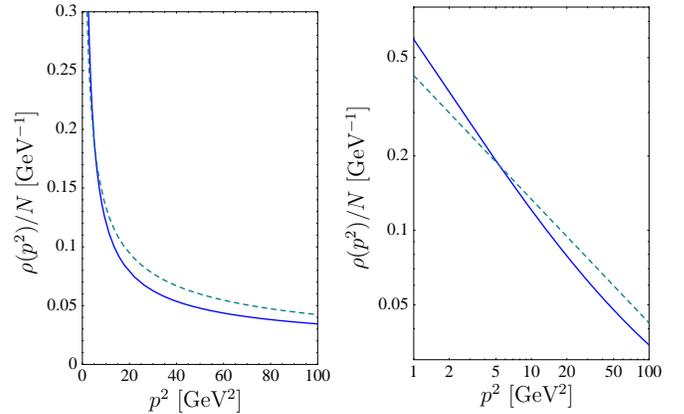}
\end{center}
\caption{\label{fig:rho}
Comparison of the unparticle spectral density (\ref{rho}) (dashed) and the spectral density (\ref{sonice}) of a massless quark jet at next-to-leading order in QCD (solid). We use parameters $M=10$\,GeV and $\eta=0.5$. The right plot shows the results on logarithmic scales.}
\end{figure}

When the tree-level approximation $\widetilde j=1$ is used in (\ref{sonice}), the result exactly coincides with the unparticle spectral density (\ref{rho}). The terms of order $\alpha_s(M)$ in $\widetilde j$ lead to logarithmic modifications of the simple power form. In the ``unparticle language" they would indicate a small breaking of conformal invariance, which as we discussed is unavoidable if the unparticle sector is coupled to the Standard Model. Therefore, our result (\ref{sonice}) shares all features of a realistic model for the spectral function of the unparticles of a conformal sector coupled to the Standard Model. In Figure~\ref{fig:rho} we compare the results (\ref{rho}) and (\ref{sonice}) for a particular set of input parameters.

In our ``interacting particle model" for unparticle states the exponent $\eta=d_{\cal U}-1$ is expressed as an integral over the cusp anomalous dimension, see (\ref{eta}). In a theory such as QCD the numerical value of $\eta$ can be O(1) provided the scales $\mu_0$ and $M$ are widely separated. This is because the perturbative smallness of the cusp anomalous dimension is overcome by the logarithmic integration over scales. In leading logarithmic approximation one finds
\begin{equation}
   \eta\approx\frac{\Gamma_0}{\beta_0}\,
   \ln\frac{\alpha_s(\mu_0)}{\alpha_s(M)} 
\end{equation}
with $\Gamma_0=4C_F$ and $\beta_0=\frac{11}{3}C_A-\frac23 n_f$. Considering the case $M=10$\,GeV as an example, we obtain $\eta=0.5$ for $\mu\approx 1.2$\,GeV. Other examples of jet functions have a similar functional form but different values of $\eta$. For the example of a gluon jet the one-loop coefficient $\Gamma_0=4C_A$ is a factor 9/4 larger than in the case of a quark jet (for $N_c=3$), leading to even larger $\eta$ values.

Why should any of the above be relevant to unparticle physics? Consider a scenario in which the Standard Model is coupled via heavy messenger exchange (with mass $\sim M_X$) to a hidden sector with a new non-abelian gauge theory we shall call QCD', which is asymptotically free at high energy but confining below some scale $\Lambda_{\rm QCD'}$. Then at energies below the messenger scale the effective interactions of Standard Model fields with the fields of the hidden sector will have the generic form (\ref{genericint}) with $\Lambda_{\cal U}$ and $M_{\cal U}$ replaced by $M_X$. In scattering or decay processes involving Standard Model fields the new massless (or light) degrees of freedom in the hidden sector can be radiated off. The energy of these new particles can range from a scale of order the characteristic energy of the process down to the confinement scale $\Lambda_{\rm QCD'}$. The jet functions in the hidden sector must then be evolved over a large energy window, and hence the resulting value of $\eta$ can be $O(1)$. Note that the situation closely resembles that for the unparticle scenario discussed below Eq.~(\ref{rhom}) in that scaling arises over a window whose upper value is determined by the energy release of the process, while its lower value is set by a dynamical scale: $\Lambda_{\rm CSB}$ in the unparticle case and $\Lambda_{\rm QCD'}$ in the QCD-inspired case. The fact that in the QCD-inspired model the scaling exponent $\eta$ itself depends on the characteristic energy of the process -- via the upper integration limit in (\ref{eta}) -- implies that the scaling behavior will be process dependent. In principle this feature could be used to distinguish the two scenarios. However, given that we do not know at present how different kinds of unparticles would couple to different kinds of Standard Model fields, the same may be true in the unparticle scenario.

The discussion of this section may be generalized to the case of massive QCD jets. If the quark field $\psi$ in (\ref{jetfun}) has mass $m$, then relations (\ref{jetfun})--(\ref{Jrge}) remain valid, but the solution (\ref{sonice}) must be modified. In this case it is no longer possible to write the solution in closed form, however a perturbative expansion of the resummed spectral function can still be obtained \cite{Boos:2005qx,inprep}. At one-loop order one finds
\begin{eqnarray}
   \rho(p^2,m^2,\mu_0) 
   &=& N(M,\mu_0)\,\frac{e^{-\gamma_E\eta}}{\Gamma(\eta)} 
    \left( p^2 - m^2 \right)^{\eta-1} \\
   &\times& \left[ 1 + \frac{C_F\alpha_s(M)}{4\pi}\,
    g\Big(\frac{p^2}{M^2},\frac{m^2}{p^2},\eta\Big) 
    + \dots \right] . \nonumber
\end{eqnarray}
The exact expression for the next-to-leading order correction (the $g$ term) is complicated and can be found in \cite{inprep}. Up to small perturbative corrections this coincides with the model density (\ref{rhom}).

Our discussion so far focused on the simplest example of a quark jet, which provides a model for a fermionic unparticle. However, the resulting spectral density (\ref{sonice}) has the same form as that for a scalar unparticle. Likewise, the obvious generalization to a gluon jet would provide a model for the spectral density of a vector unparticle.

\section{Conclusions}

The proposal of unparticle degrees of freedom as effective low-energy fields describing the interactions of Standard Model particles with a hypothetical, yet unexplored conformal sector \cite{Georgi:2007ek} has opened a playground for theoretical speculations about possible signatures in present and future experiments \cite{Georgi:2007si}--\cite{Delgado:2007dx}. It has been argued that the characteristic dependence of the unparticle spectral density on momentum, as reflected by the fractional-power behavior in (\ref{rho}), could serve as a ``smoking gun" signature of conformal invariance in the hidden sector. 

In this Letter we have shown that the resummation of Sudakov logarithms for the propagators of fermions and gauge bosons in interacting theories such as QCD produces spectral densities for massless and massive particles that are virtually indistinguishable from those of unparticles. The differences are of the form of small logarithmic corrections at higher orders in perturbation theory. They mimic conformal symmetry-breaking terms, which are unavoidable in all models where the unparticle sector is coupled to the Standard Model, so that it can have observable effects. 

It follows from our discussion that the degrees of freddom which Georgi has called ``unparticle stuff" do, indeed, very much behave like ordinary interacting particles. This does not mean that unparticle physics is uninteresting, as behavior such as (\ref{rho}) may indeed arise from a conformal (or nearly conformal) sector weakly coupled to the Standard Model. However, our analysis shows that unparticle signatures are less striking than originally advocated. Very similar effects can arise in models where the hidden-sector theory resembles a theory such as QCD. 

\vspace{0.3cm}  
{\em Acknowledgments:\/} 
My interest in this subject was stimulated by a talk delivered by John Terning at the E\"otv\"os-Cornell workshop held in Budapest in June 2007. Fermilab is operated by Fermi Research Alliance, LLC under contract with the U.S.\ Department of Energy.

\end{document}